# Unconditionally secure commitment in position-based quantum cryptography

Muhammad Nadeem
Department of Basic Sciences,
School of Electrical Engineering and Computer Science
National University of Sciences and Technology (NUST)
H-12 Islamabad, Pakistan
muhammad.nadeem@seecs.edu.pk

A new commitment scheme based on position-verification and non-local quantum correlations is presented here for the first time in literature. The only credential for unconditional security is the position of committer and non-local correlations generated; neither receiver has any pre-shared data with the committer nor does receiver require trusted and authenticated quantum/classical channels between him and the committer. In the proposed scheme, receiver trusts the commitment only if the scheme itself verifies position of the committer and validates her commitment through non-local quantum correlations in a single round. The position-based commitment scheme bounds committer to reveal valid commitment within allocated time and guarantees that the receiver will not be able to get information about commitment unless committer reveals. The scheme works for the commitment of both bits and qubits and is equally secure against committer/receiver as well as against any third party who may have interests in destroying the commitment. Our proposed scheme is unconditionally secure in general and evades Mayers and Lo-Chau attacks in particular.

## 1. Introduction

A bit commitment scheme involves two mistrustful parties, committer and a receiver, conventionally called Alice and Bob. Alice provides a piece of evidence to Bob that commit her to a specific bit b. At this stage, Bob should not be able to extract the bit value from that evidence. At a later time, however, it must be possible for Bob to know the genuine bit value b with absolute guarantee when Alice reveals the committed bit and Alice should not be able to change her mind about the value of the bit b.

Bit commitment is an essential cryptographic protocol for implementing a wide range of other functions. From a trusted bit commitment scheme one can construct coin tossing[1], digital signature[2], oblivious transfer[3], and hence two-party secure computation[4]. But, unfortunately, all classical bit commitment schemes based on unproven computational hardness cannot be secure in principle.

Many authors attempted to achieve information-theoretically secure bit commitment in non-relativistic quantum cryptographic settings[5-7]. Currently, however, it is known that all bit commitment schemes based on non-relativistic quantum information cannot evade Mayers and Lo-Chau attacks[8-11] if Alice and Bob do not pre-share any data. Mayers showed that bit commitment protocols that make use of the theory of special relativity are also insecure[8]. Later on, A. Kent proposed relativist classical bit commitment protocols[12,13] and showed that security against all classical attacks as well as quantum attacks introduced by Mayers and Lo-Chau is guaranteed by impossibility of superluminal signaling. However, he has not given any proof of unconditional security of these protocols against general attacks. A. Kent also proposed a general framework for relativistic cryptographic protocols where he considered that the background space time is approximately Minkowski and communicating agencies have distributed agents throughout the space time. If a sender sends a state from point ($x$,0) by performing instantaneous computation on it then after some fixed time t > 0, the light-like separated agents from the sender in some given inertial frame can receive the state on a special sphere of radius t and centered at $x$. The assumption of instantaneous computations in a fixed inertial frame breaks the Lorentz invariance which is necessary for practical cryptographic tasks.

Recently, A. Kent addressed the same problem in relativistic quantum cryptography[14,15]. Both of his protocols are based on similar intuition and are secure against general classical and quantum attacks[16,17]. The security of these protocols depends on the no-cloning[18] and no-signaling principles. Moreover, two different groups experimentally demonstrated the results of Kent's protocol[15], using quantum communication and special relativity, with commitment times of 15ms[19] and 30μs[20], respectively.

Here, they used the idea of disjoint sites for both Alice and Bob first introduced by Ben-Or *et al*[21]. In this procedure both Alice and Bob are not the individuals but they have their own trusted agents (≥ 2) distributed throughout the space time. Both Alice and Bob need to have secure and authenticated quantum/classical channels to communicate with their respective agents. Moreover, these protocols require that both Alice and Bob agree in advance on some space time point where the commitment will commence and both these parties must have disjoint laboratories near that point. The receiver Bob sends the encrypted quantum states to the committer Alice at that point and then Alice sends these quantum states (or measurement results) after processing to her space-like separated agents in causal future of that point. The committed bit value by Alice then depends on the direction in which she sends the information. To reveal the committed bit, Alice's agent in that direction sends the information to the Bob's agent in nearby laboratory. The minimum secure laboratories required by these protocols are six: three in possession of Alice and her two agents while three kept by Bob and his two agents.

In this article, we present a commitment scheme based on position-verification and non-local quantum correlations, different from all previously proposed schemes in literature. Position-based quantum cryptography[22-33] has remained a conundrum for many years, but recently we proposed a quantum scheme for secure positioning[34] that can be useful in implementing other cryptographic tasks[35]. This strategy for studying cryptography uses teleportation[36] and entanglement swapping[37] in an interesting and new fashion.

Our procedure for position-based commitment has remarkable advantages over previously presented schemes. Our scheme bounds Alice to commit while she initiates the scheme instead of sending signals in different directions to her agents after processing data. In fact, Alice does not need any single agent while Bob needs only one trusted agent who helps him in position-verification of Alice. Moreover, Alice and Bob do not need to agree on a pre-defined position to start commitment. Instead, Alice and Bob can be anywhere in Minkowski space time. Alice starts the commitment scheme and Bob can only trust her commitment if the commitment scheme itself verifies position of Alice and validates the commitment through non-local quantum correlations. Moreover, our scheme is equally secure against committer/receiver as well as against any third party (eavesdroppers) who may have interests in destroying this commitment. Furthermore, our scheme works for the commitment of both bits and qubits and does not require any advanced technology (quantum computer) for the commitment of quantum superposition states $(|0\rangle \pm |1\rangle)/\sqrt{2}$. Finally, our scheme is purely relativistic quantum mechanical and does not require any secure classical channel, different from previously proposed schemes. All the classical information can be communicated (or can be publically announced) after completion of quantum transmissions, so even insecure classical channels do not lead to cheating.

## 2. Proposed commitment scheme

We assume that committer Alice, receiver Bob and his agent have fixed position in Minkowski space-time; their secure sites. Bob and his agent have précised and synchronized clocks also. We also assume that Bob and his agent are trusted and known to each other, they need to share maximally entangled states among them. They can pre-distribute these states or send entangled qubits over quantum channels encrypted by pre-shared classical keys. However, all the quantum/classical channels between Alice and Bob are insecure and they have no pre-shared quantum/classical data. In short, Alice and Bob have neither met before nor share any secret information and cannot trust anything happening outside their secure laboratories. Finally, we suppose that all quantum/classical signals can be sent at the speed of light while the time for information processing at the sites of Alice and Bob is negligible. For simplicity, we assume that Alice, Bob and his agent are collinear such that Alice is at a distance d from both Bob and his agent. Our scheme is described below where we denote Alice by A while Bob and his agent by B and C respectively and we write Bell states as:

$$|u_i u_j\rangle = \frac{|0\rangle|u_j\rangle + (-1)^{u_i}|1\rangle|1 \oplus u_j\rangle}{\sqrt{2}} \quad (1)$$

where $u_i u_j \in \{0,1\}$ and $\oplus$ denotes addition with mod 2.

Commitment phase: The receiver B and his agent C share randomly chosen N maximally entangled qubit pairs, $|\beta_{bc}\rangle = \otimes_{n=1}^{N} |u_b u_c\rangle_n$ where $|u_b u_c\rangle_n$ are labeled Bell states. To make a commitment, A secretly prepares N Bell pairs $|\beta_{ac}\rangle = \otimes_{n=1}^{N} |u_a u_c\rangle_n$ and sends one qubit of each pair, $|u_c\rangle_n$, to C through insecure quantum channel. A also sends a dummy quantum state $|d\rangle$ to B simultaneously and announces her position. This dummy state carries no secret information but just informs B that A has made her commitment. Here A and B agree on a code: To commit to the bit 0, A initiates the scheme with Bell pairs $|\beta_{ac}\rangle = \otimes_{n=1}^{N} |00\rangle_n$; to commit to the bit 1, A initiates with $|\beta_{ac}\rangle = \otimes_{n=1}^{N} |01\rangle_n$; to commit to the qubit $(|0\rangle + |1\rangle)/\sqrt{2}$, A prepares $|\beta_{ac}\rangle = \otimes_{n=1}^{N} |10\rangle_n$ while for committing $(|0\rangle - |1\rangle)/\sqrt{2}$, A starts the scheme with Bell pairs $|\beta_{ac}\rangle = \otimes_{n=1}^{N} |11\rangle_n$.

At time t, when C (B) receives entangled qubits (dummy state) from A, C applies randomly chosen unitary operators from set $\{\sigma_x, \sigma_z, \sigma_z \sigma_x\}$ on these qubits. He then performs BSM[38] on qubits in his possession (N/2 entangled with B and N/2 with A) and gets two classical bits for each pair. Let's label BSM results of C as classical 2N-bit string $\beta_{cc'} = \otimes_{n=1}^{N} (u_c u_{c'})_n$. This measurement of C projects the qubits in possession of A and B into Bell states $|\beta_{ss'}\rangle = \otimes_{n=1}^{N} |u_s u_{s'}\rangle$ instantly, unknown to everyone at the moment[33].

Simultaneously, at time t, B prepares N qubits in the state $|\phi\rangle = \otimes_{n=1}^{N} |u_i\rangle_n$ where $u_i \in \{0,1\}$, applies random unitary operator $U_n$ on each qubit and teleports the state $|\psi\rangle = \otimes_{n=1}^{N} U_n |u_i\rangle_n$ to A. If BSM result of B is $u_b u_{b'}$ while teleporting $n^{th}$ qubit, then A's half of the $n^{th}$ Bell's state will become one of the corresponding four possible states $|\psi'\rangle_n = \sigma_z^{k_n} \sigma_x^{k'_n} U_n |u_i\rangle_n$. Here $k_n$ and $k'_n$ depend on which one of the four entangled state is shared between A and B at time of teleportation. For example, if they share Bell state $|00\rangle$ then $k_n = u_b$ and $k'_n = u_{b'}$. If shared state is $|01\rangle$ then $k_n = u_b$ and $k'_n = 1 \oplus u_{b'}$. If they share $|10\rangle$ then $k_n = 1 \oplus u_b$ and $k'_n = u_{b'}$ while for $|11\rangle$, $k_n = 1 \oplus u_b$ and $k'_n = 1 \oplus u_{b'}$. Again $|\psi'\rangle_n$ is unknown to everyone. Since information processing time at the sites of A and B is negligibly small, so we are not considering it.

Revealing phase: At the same time t, A applies further unitary operators $\sigma_z^{1 \oplus (u_a)_n} \sigma_x^{(u_c)_n}$ on $|\psi'\rangle_n$ and immediately returns the state

$$|\psi''\rangle = \otimes_{n=1}^{N} \sigma_z^{1 \oplus (u_a)_n} \sigma_x^{(u_c)_n} \sigma_z^{k_n} \sigma_x^{k'_n} U_n |u_i\rangle_n \quad (2)$$

to B over insecure quantum channel between them. Simultaneously she reveals her commitment by telling values of $u_a u_c$, corresponding to her initially prepared Bell pairs, to both B and C.

Suppose both B and C receives this information at time T. C also sends time t, T, applied unitary operators and his BSM measurement results $\beta_{cc'} = \otimes_{n=1}^{N} (u_c u_{c'})_n$ to B. Now B can find the exact swapped entangled states between him and A, $|\beta_{ss'}\rangle = \otimes_{n=1}^{N} |u_s u_{s'}\rangle$, and (hence) the exact values of $k_n$ and $k'_n$. He applies corresponding unitary transformations on $|\psi''\rangle$ and gets the result. If B gets verified state $|\phi\rangle = \otimes_{n=1}^{N} |u_i\rangle_n$ and receives information from A within allocated time, T-t = d/c, then he is certain about the position of A and her commitment.

### 3. Security analysis; N=3

In a commitment scheme, the committer A can cheat by choosing a particular bit value b (or qubit value q) during the commitment phase and tell the receiver B another value during the revealing phase. A commitment scheme is said to be secure against A only if such a fake commitment can be discovered by B in the revealing phase.

Let's first consider a simple case for N=3 where A is honest. In the committing phase, suppose A starts with the state $|\beta_{ac}\rangle = |01\rangle_1 |01\rangle_2 |01\rangle_3$ to commit to the bit value 1. Now if C applies randomly chosen operators $\{(\sigma_z \sigma_x)_1, (\sigma_z)_2, (\sigma_x)_3\}$ on qubits received from A, then the entangled states shared between A and C will become $|\beta'_{ac}\rangle = |10\rangle_1 |11\rangle_2 |00\rangle_3$, unknown to everyone now. If B and his agent C share randomly prepared entangled state $|\beta_{bc}\rangle = |01\rangle_1 |11\rangle_2 |10\rangle_3$ and C gets BSM results $\beta_{cc'} = (00)_1 (10)_2 (11)_3$ then swapped entangled state between A and B will be $|\beta_{ss'}\rangle = |11\rangle_1 |10\rangle_2 |01\rangle_3$, unknown to everyone[33]. Now in the revealing phase, if A announces exact Bell states $|\beta_{ac}\rangle = |01\rangle_1 |01\rangle_2 |01\rangle_3$ (her committed bit 1) and C announces his applied unitary operators and BSM results, then B can get exact swapped state $|\beta_{ss'}\rangle = |11\rangle_1 |10\rangle_2 |01\rangle_3$, and so the values of

$$[k_1 k'_1][k_2 k'_2][k_3 k'_3] = [(1 \oplus u_b)(1 \oplus u_{b'})][(1 \oplus u_b) u_{b'}][u_b (1 \oplus u_{b'})] \tag{3}$$

He then gets the verified state $|\phi\rangle$.

Now if the committer A is not honest and tries cheating strategies introduced by Mayers and Lo-Chau, we show here that our scheme is unconditionally secure with 100% success rate. Let's suppose A changes her mind and announces $|\beta_{ac}\rangle = |11\rangle_1 |11\rangle_2 |11\rangle_3$, fake commitment, and tries to convince B by applying random unitary operator $\{(\sigma_z)_1, (\sigma_x)_2, (\sigma_z \sigma_x)_3\}$ on her retained qubits. She will do this on hit and trail basis as she does not know either BSM results of C and his operations on $|\beta_{ac}\rangle$ or initially shared entangled states between B and C. Hence, entangled state swapped between A and B will become $|\beta'_{ss'}\rangle = |01\rangle_1 |11\rangle_2 |10\rangle_3$. Now B receives fake commitment from A, $|\beta_{ac}\rangle = |11\rangle_1 |11\rangle_2 |11\rangle_3$, while $\{(\sigma_z \sigma_x)_1, (\sigma_z)_2, (\sigma_x)_3\}$ and $\beta_{cc'} = (00)_1 (10)_2 (11)_3$ from C. He will extract Bell state shared between A and C as $|\beta''_{ac}\rangle = |00\rangle_1 |01\rangle_2 |10\rangle_3$ and (hence) the swapped state between him and A as $|\beta''_{ss'}\rangle = |01\rangle_1 |00\rangle_2 |11\rangle_3$, that are different form $|\beta'_{ss'}\rangle = |01\rangle_1 |11\rangle_2 |10\rangle_3$ and (hence) wrong. Finally, B will calculate wrong values of $[k_1 k'_1][k_2 k'_2][k_3 k'_3]$, and hence wrong state $|\phi\rangle$. In conclusion, if A tries to cheat by applying unitary operations on retained qubit, B will always deduce wrong values of $k_n$ and $k'_n$ and will not get initially prepared state $|\phi\rangle$ and (hence) discover the cheating with 100% success. That is, B can only receive exact state back and accepts A's commitment if and only if A remains fair.

Furthermore, Alice cannot cheat successfully if she delays announcements of her initially prepared Bell states indefinitely and waits to get handful information for cheating. In such a situation, Bob will reject the commitment instantly as he will not get the response within allocated time. In conclusion, position-based quantum commitment based on non-local quantum correlations forces Alice to reveal valid commitment within time.

Similarly, the scheme is equally secure from Bob. What do we mean by secure against Bob? In a commitment scheme, Bob can cheat by deducing the committed bit before the revealing phase. In our proposed commitment scheme, Bob can cheat only if he knows BSM results of his agent and the swapped entangled states between him and A before the revealing phase. Even if he manages somehow, by non-local measurements for example, to get BSM result from his agent, he cannot find swapped states between him and A. Hence, he will remain unable to get information about commitment unless Alice reveals.

Furthermore, we suppose that committer and receiver are distant parties and do not have any pre-shared data or secure quantum/classical channels between them. In such a situation, eavesdroppers can try to destroy the scheme even if both committer and receiver are fair. We would like to mention that our scheme is secure against any group of adversaries having unlimited pre-shared entangled states[34]. In conclusion, our proposed position-based commitment scheme is unconditionally secure against receiver, committer and any group of eavesdroppers who have infinite amount of pre-shared entanglement and power of non-local quantum measurements in negligible time.

**4. Discussion**
We presented a commitment scheme based on quantum position-verification and non-local correlations. Our scheme works for the commitment of both bits and qubits and it does not require any advance quantum computer for committing quantum superposition states (+ and -). Remarkable feature of this scheme is that it does not imply any constraint on either receiver or committer. They need not fix either committing position or revealing time priorly but rather the scheme itself verifies these data.

Our scheme assumes that Bob and his agent have synchronized clocks. This assumption of synchronization simplifies the scheme but does not give any advantages to eavesdroppers to cheat. In the scheme, Bob and his agent are trusted and have authenticated channels between them. Bob's agent can simply send his/her position to Bob with one-time pad so distance 2d between them is known to both of them. Now they can synchronize their clocks by sending light signals and computing round trip times. So first of all, it is impossible for eavesdropper to desynchronize clocks of Bob and his agent and hence cannot cheat. Secondly, even under week assumption where Bob and his agent have not synchronized clocks, they can note the communication times with Alice independently and then compare the results that must be identical; T-t=T'-t'=d/c.

We would also like to mention here that our scheme assumes fixed space time positions of Alice, Bob and Bob's agent. The proposed Position-based commitment scheme adds an additional layer of security where position of Alice is used as her credential along with non-local quantum correlations between Alice and Bob. If Alice makes a commitment at position P (distance d away from both Bob and his agent) in the commitment phase and later tries to cheat while revealing commitment from different position P', it would not help her at all. She will have to respond in time d/c and within this time she cannot get any useful information about non-local correlations generated to cheat.

The scheme can be generalized where receiver has multiple agents distributed in Minkowski space time and committer communicates with them through multi-particle entangled states over insecure quantum channels. It will allow the committer to commit any arbitrary quantum superposition state $|\psi\rangle = \alpha|0\rangle + \beta|1\rangle$ also.

Our proposed scheme can be efficiently and reliably implemented using existing quantum technologies and its security relies on physical principles, no-cloning and no-signaling, instead of computational hardness. We hope the scheme presented here will lead to further theoretical and experimental investigation of bit/qubit commitment. In the broader perspective, proposed commitment scheme can be useful to investigate other cryptographic tasks like coin tossing, digital signature and secure two-party computation, in a secure and robust way. In conclusion, this commitment scheme based on our recently proposed quantum position-verification and non-local quantum correlations will imply a great excitement in the area of position-based quantum cryptography.